\documentclass{aa}
\usepackage{psfig}
\newcommand{\be}{\begin{equation}}
\newcommand{\en}{\end{equation}}
	
\def\zabs{$z_{\rm abs}$}
\def\zem{$z_{\rm em}$~}
\def\lya{Ly$\alpha$ }

\def\h2{H$_2$}
\def\hi{H~{\sc i}~}

\def\nv{N~{\sc v}~}

\def\cii{C~{\sc ii}~}

\def\civ{C~{\sc iv}~}

\def\nva{N~{\sc v}$\lambda$1238~}
\def\nvb{N~{\sc v}$\lambda$1242~}
\def\civab{C~{\sc iv}$\lambda\lambda$1548,1550~}

\def\siii{Si~{\sc ii}~}

\def\siiv{Si~{\sc iv}~}

\def\siivb{Si~{\sc iv}$\lambda$1402~}
\def\siivab{Si~{\sc iv}$\lambda\lambda$1393,1402~}

\def\kms{km~s$^{-1}$}
\begin{document}
%
\title{
Outflowing material in the \zem = 4.92 BAL QSO
 SDSS~J160501.21$-$011220.0\thanks{Based on observations carried out at the
 European Southern Observatory (ESO) under programmes 67.A-0078 and 69.A-0457
 with the UVES spectrograph installed at the Nasmyth focus B of the VLT 8.2m
 telescope,unit Kueyen,on Cerro Paranal in Chile}
}
\titlerunning{ BAL QSO SDSS J160501.21-011220.0}  
\author{Neeraj Gupta\inst{1}, Raghunathan Srianand\inst{2},  
Patrick Petitjean\inst{3,4} \& C\'edric Ledoux\inst{5} }
\authorrunning{Neeraj et al}
\institute{
$^1$NCRA, Post Bag 3,  Ganeshkhind, Pune 411 007, India\\
~\email{neeraj@ncra.tifr.res.in}\\
$^2$IUCAA, Post Bag 4, Ganeshkhind, Pune 411 007, India\\
~\email{anand@iucaa.ernet.in}\\ 
$^3$Institut d'Astrophysique de Paris -- CNRS, 98bis Boulevard 
Arago, F-75014 Paris, France\\
$^4$LERMA, Observatoire de Paris, 61 Rue de l'Observatoire,
   F-75014 Paris, France\\
~\email{petitjean@iap.fr}\\
$^5$European Southern Observatory, Alonso de C\'ordova 3107, Casilla 19001,
Vitacura, Santiago, Chile\\ 
~\email{cledoux@eso.org}\\
}
\date{Received date/ Accepted date}
\offprints{R. Srianand}

\abstract{
We present the analysis of broad absorption lines (BALs) seen in the
spectrum of the \zem $\simeq$4.92 QSO SDSS J160501.21-011220.0. 
Our high spectral resolution UVES spectrum shows two well detached 
absorption line systems at \zabs= 4.685 and 4.855. The system at
\zabs= 4.855 covers the background source completely suggesting 
that the gas is located outside the broad emission line region. 
On the contrary the system at \zabs= 4.685, which covers only on the continuum
source, has a covering factor of the order of 0.9.
Physical conditions are investigated in the BAL system 
at \zabs= 4.855 using detailed photoionization models.
The observed \hi absorption line together with the limits on \cii and \siii 
absorptions suggest that 16~$<$~log $N$(H~{\sc i}) (cm$^{-2}$)~$<$~17 in this system.
Comparison with models show that the observed column densities of 
\nv, \siiv and \civ in this system require that 
nitrogen is underabundant by more than a factor 3 compared to 
silicon if the ionizing radiation is similar to a typical 
QSO spectrum. This is contrary to what is usually derived for the
emission line gas in QSOs.
We show that the relative suppression in the \nv column density can be 
explained for Solar abundance ratios or abundance ratios typical 
of Starburst abundances if an ionizing spectrum devoid of X-rays is 
used instead. Thus, if the composition of  BAL is 
like that of
the emission line regions it is most likely that the cloud sees
a spectrum devoid of X-rays similar to what we observe from this QSO. This is 
consistent with the fact that none of our models have high 
Compton optical depth to remove X-rays from the QSO.
Similar arguments lead to the conclusion that the system at \zabs = 4.685
as well is not Compton thick. 
Using simple Eddington arguments we show that the mass of the central
black hole is $\sim 8\times 10^8$ M$_\odot$. This suggests that the 
accretion onto a seed black hole must have started as early as
$z$ = 11. 
}
\maketitle
\begin{figure*}
\centerline{\vbox{
\psfig{figure=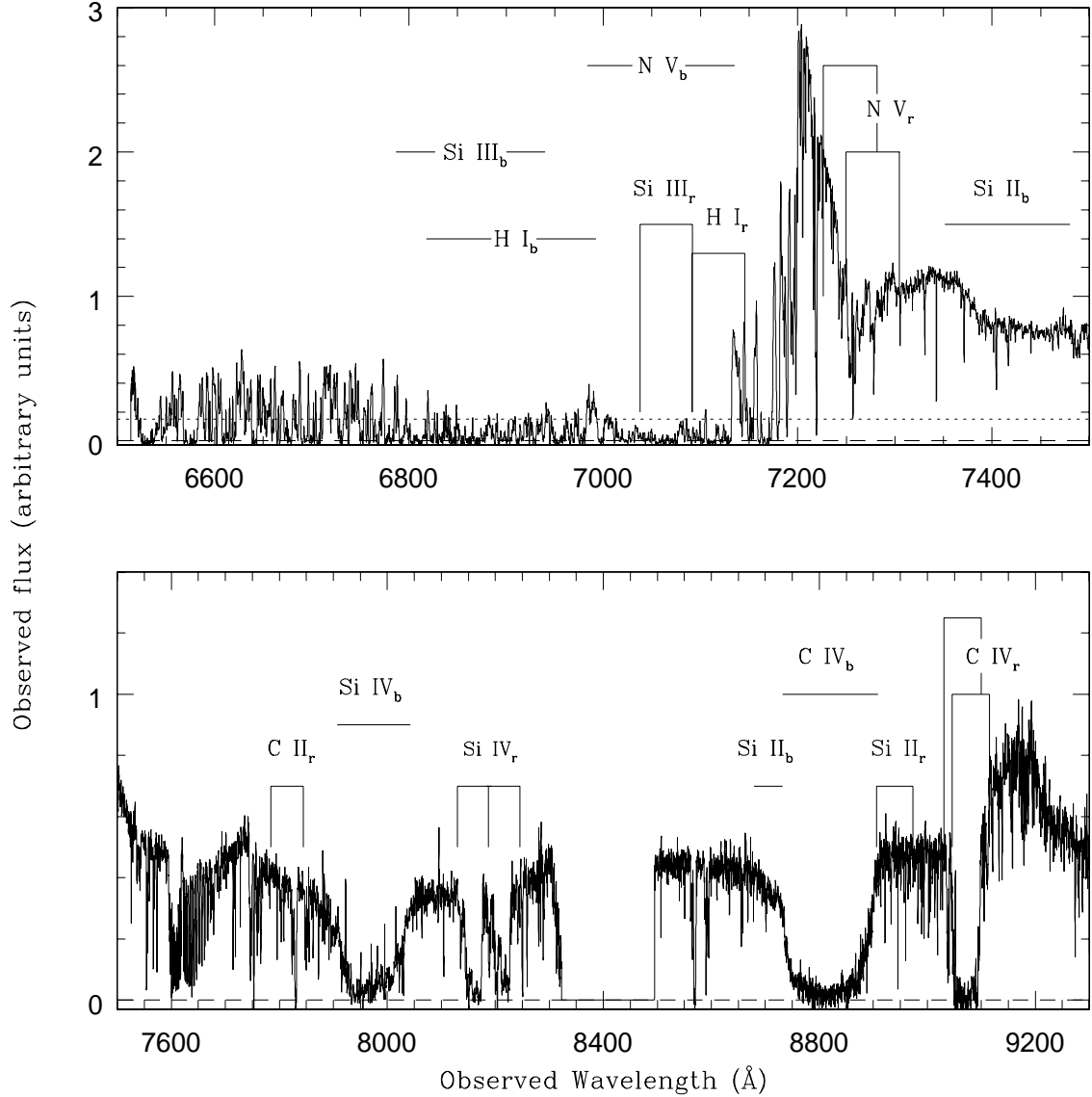,height=16.5cm,width=16.0cm,angle=360}
}}
\caption[]{UVES spectrum of the \zem = 4.92 QSO SDSS J160501.21-011220.
Two well detached BAL features are seen covering the redshift ranges
4.62$-$4.75 and 4.83$-$4.88. We denote them as blue and red systems
respectively. The expected positions of different lines (with
sub-scripts $r$ and $b$ denoting red and blue respectively) are marked with
solid lines. The horizontal dotted line in the top panel gives the 
mean transmitted flux computed over the \lya forest in the wavelength
range 6550$-$6800 \AA~ that is not affected by the BAL absorption lines.
}

\label{fig1}
\end{figure*}
\section{Introduction}
Broad absorption line systems (BALs) seen in the spectra of QSOs are
characterized by absorption features with large velocity widths 
( $\Delta v \sim$ a few $1000$ \kms) 
and, usually, high ionization states (Turnshek et al. 1988; Weymann et al. 1991).
The gas giving rise to  BALs is believed to be material ejected
by the quasar but still located very close to the central regions.
Thus studying metallicities in the absorbing gas is a direct probe of the 
chemical enrichment in the very central regions of AGNs 
(see Hamann \& Ferland 1999 for the review). The associated absorption line
systems by and large show large metallicities with a tendency
of nitrogen being over-abundant with respect to oxygen and other 
$\alpha-$ process elements (e.g. Petitjean et al. 1994; Korista et al. 
1996; Hamann 1997; Petitjean \& Srianand 1999). 
Detailed studies of emission line
properties of $z\ge 4$ QSOs have revealed Super-Solar metallicities in
the line emitting gas in broad line regions (BLR), suggesting 
very rapid star formation process (e.g. Hamann \& Ferland 1992;
Dietrich et al. 2003).
However, no such metallicity estimates at such  high redshifts based 
on absorption lines using echelle spectra are available.

\par\noindent
Unlike in the case of 
most normal galaxies, it is believed that the high metal enrichment usually
observed in QSOs has taken place in less than a few 10$^8$ years 
through rapid star formation similar to what is expected 
to happen in the center of ellipticals (Hamann \& Ferland 1993; 
Matteucci \& Padovani 1993). Thus, studying BAL QSOs at \zem$ \ge 5$, 
where the age of the Universe is t$\le 1.07\times 10^9$~yr (for H$_{\rm o}$ = 
75 km~s$^{-1}$ Mpc$^{-1}$, $\Omega_\Lambda$ = 0.70, $\Omega_{\rm m}$ = 0.30), 
is important for understanding the star-formation history at a time close to the 
epoch of reionization and possibly also for constraining cosmological parameters.
In the initial list of SDSS QSOs (Fan et al. 2000) there are few QSOs
at $z\ge 4$ with associated
absorption lines. We obtained a UVES spectrum of the QSO SDSS J160501.21-011220.0,
the highest redshift BAL QSO known at that time, with the aim to study
the outflowing gas in detail.   

\par\noindent
In order to derive a realistic estimate of the 
absorbing gas metallicity one needs to have a good handle on the ionization
corrections. Associated systems being close to the QSO are most
probably ionized by the QSO light rather than the diffuse intergalactic
background. However the ionizing spectrum from the BAL QSOs is 
poorly known.  It is known that BAL QSOs are under-luminous in X-rays (Bregman
1984; Singh et al. 1987; Green \& Mathur 1996). Recent, deep
Chandra observations have shown that the optical to X-ray spectral
index, $\alpha_{OX}$, measured for BAL QSOs is systematically lower than
that of non-BAL QSOs (Green et al. 2001). It is not clear whether
this X-ray weakness is due to the QSO being intrinsically 
X-ray quiet (i.e., this is a property of the central engine) or 
due to line-of-sight absorption (i.e., radiative transfer effects). Using a complete 
sample of optically selected QSOs, Brandt et al. (2000) have shown a significant
anti-correlation between $\alpha_{OX}$ and the total equivalent width
of the \civ absorption lines. In addition it is noticed that the 
emission line properties of the BAL and non-BAL QSOs are {\bf very similar}
even though the observed X-ray properties differ significantly
(Weymann et al. 1991; Korista et al. 1993). Thus, it is most 
likely that the relative X-ray weakness of BAL QSOs is due to 
intrinsic absorption. 

\par\noindent
It must be noted that all existing models of BAL systems assume
standard QSO spectrum. However estimated metallicities will be 
different if the intrinsic $\alpha_{OX}$ is different from that
of a typical QSO spectrum. Indeed, Srianand \& Petitjean (2000)
have shown that the inferred metallicities decrease with decreasing
value of $\alpha_{OX}$.
Also, if the absorption is responsible 
for the suppression of X-rays then it is important to investigate whether it is caused
by the gas responsible for the BAL troughs (common absorbers) or by a distinct gas
component.  It is likely that the common absorber picture is correct
for the BAL towards PHL 5200 (Mathur et al. 1995) and that a distinct Compton
thick screen is required for the BAL towards PG~0946+301 (Gallagher et al.
1999; Mathur et al. 2000; see also Kraemer et al. 2002).  

In this paper we present a detailed analysis of two well detached BAL 
outflow components that are seen in the spectrum of QSO J160501.21-011220.0  
($z_{em}$~=~4.92). 
Details of the observations are discussed in Sect.~2 which is
followed by the description of the BALs and column density estimates
in Sect.~3. In Sect.~4 we investigate the physical conditions of the
gas.The results are summarised in Sect.~5.

\section{Observations and data reduction}
The Ultraviolet and Visible Echelle Spectrograph (UVES; Dekker et al. 
2000), installed at the ESO VLT 8.2 m telescope, unit Kueyen, on Mount 
Paranal in Chile, was used on June 15-17, 2001, to obtain high resolution 
spectra of QSO J\,160501.21$-$011220.0.
A non-standard setting with cross-disperser \#4 and central wavelength
8420 \AA\ was used in the Red arm of UVES. Full wavelength coverage was
obtained this way from 6514 to 8300 \AA\ and from 8494 to 9300 \AA\
accounting for the gap between the two Red-arm CCDs. The CCD pixels 
were binned $2\times 3$ (namely, twice in the spatial direction and three 
times in the dispersion direction) and the slit width was either fixed 
to $1\farcs 2$ or $1\farcs 5$, yielding an overall spectral resolution 
$R\sim 30000$. The total integration time 3h30min was split into 3 
exposures. The data were reduced in the dedicated context of MIDAS
\footnote{MIDAS:~Munich Image Data Analysis System, trademark of the European Southern 
Observatory (ESO)}, 
the ESO data reduction system, using the UVES pipeline 
(Ballester et al. 2000) in an interactive
mode. The main characteristics of the pipeline is to perform a precise
inter-order background subtraction for science frames and
master flat-fields, and an optimal extraction of the object signal rejecting
a number of cosmic ray impacts and subtracting the sky
spectrum simultaneously. The pipeline products are checked step by step.
The wavelength scale of the spectra reduced by the pipeline was then
converted to vacuum-heliocentric values and individual 1-D exposures
scaled, weighted and combined altogether using the NOAO {\it onedspec} package
of the IRAF\footnote{IRAF:~The Image Reduction and Analysis Facility is distributed
by the National Optical Astronomy Observatories, which is operated by the Association
of Universities for Research in astronomy,~Inc. (AURA), under cooperative agreement 
with the National Science Foundation.} software. 
During this process, the spectra were rebinned to
0.08 \AA\ pix$^{-1}$. In order to increase the signal-to-noise ratio in
this near-IR part of the optical range (and as the lines of interest are
broad), we applied a Gaussian
filter smoothing with two pixel FWHM and get an effective spectral 
resolution of $\sim 12$ km s$^{-1}$. The resulting signal-to-noise 
ratio per pixel is of the order of 20 or more over most of the 
wavelength range considered here. 
The final  spectrum is shown in Fig.~\ref{fig1}.

\par\noindent
The continuum fitting in the red side of the \lya emission
has been performed using a smooth low order polynomial considering only the 
absorption free regions. In the Lyman-$\alpha$ forest, we approximated
the QSO spectrum by a powerlaw, $f_\nu\propto\nu^{-0.6}$,
that we fitted on the low-dispersion SDSS spectrum made available to us 
by Dr. Fan.
The normalization of the power-law was done using the flux in the region 
between the \civ and \siiv emission lines. 
In our normalised spectrum the mean \lya transmission flux in the
redshift range $z$=4.39 to $z$=4.60 (that is not contaminated by 
any BAL absorption lines) is 0.15. This is consistent with
the values derived in the same redshift range using non-BAL QSOs
(Becker et al. 2001).

\section {Broad absorption line systems}
Based on the unabsorbed \civ emission line we estimate the emission 
redshift of the QSO to be \zem=4.92. Two well detached absorption line systems are 
identified based on the \civ and \siiv absorptions over the redshift
ranges, $z_{\rm abs}$=4.83$-$4.88 and $z_{\rm abs}$ = 4.62$-$4.75, 
respectively. 
For mean redshifts \zabs~= 4.855 and 4.685, 
this corresponds to ejection velocities of $\approx$3,330 \kms and 
$\approx$ 12,160 \kms. The spectrum in the Lyman-$\alpha$ forest is 
heavily affected by the high \lya opacity of
the intergalactic medium. Even though excess absorption 
is seen at the expected positions of BAL 
absorption lines (see Fig.~\ref{fig1}) it is difficult to 
estimate column densities in this region.
Therefore, in the following sections we use only lines observed on the red
side of the \lya emission line to investigate the physical
conditions in the outflowing gas.
\begin{figure}
\centerline{\vbox{
\psfig{figure=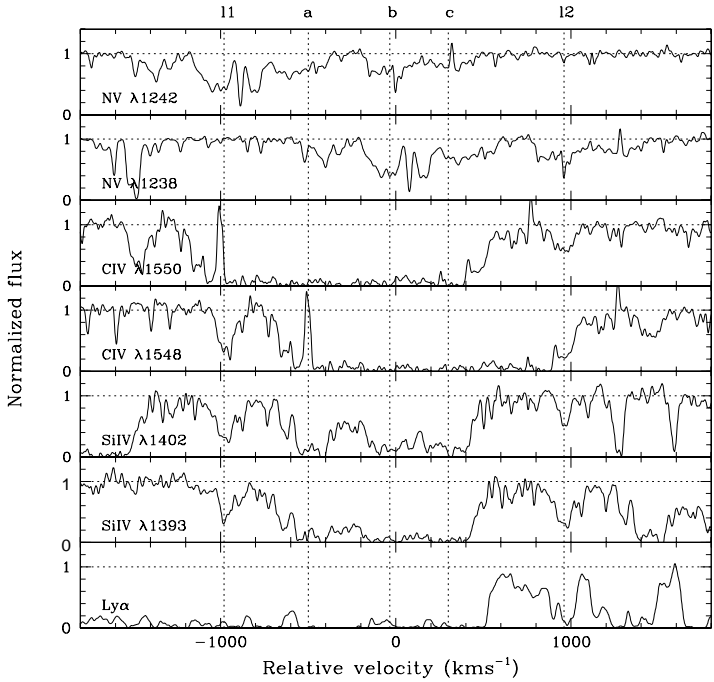,height=9.5cm,width=10.0cm,angle=0}
}}
\caption[]{Velocity plots of the red absorption line system 
centered at \zabs~= 4.855. We identify 5 distinct components
based on the \siiv and \civ profiles (marked with vertical dotted lines). 
Detection of \lya line in this system is unambiguous as the profile just 
follows the well detached \siiv absorption line at least from 
0 to 600~\kms.}
\label{fig2}
\end{figure}

\begin{figure}
\centerline{\vbox{
\psfig{figure=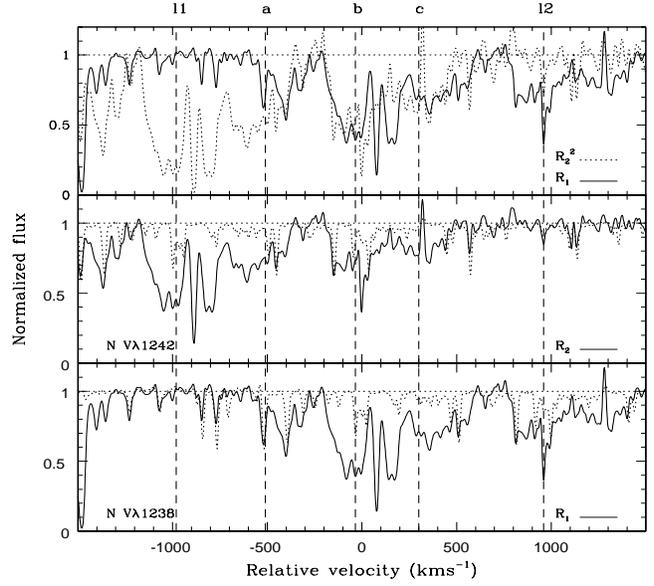,height=8.0cm,width=9.0cm,angle=0}
}}
\caption[]{
Velocity plots of \nv doublets centered at  \zabs~= 4.855. R$_1$ and R$_2$ are 
the residual intensities in the first and second lines of the doublet.Vertical 
dotted lines show the subcomponents l1, a, b, c and l2 respectively.
In the lower two  panels the dotted profiles give the atmospheric
lines seen toward QSO 1122$-$1628. In the top panel we overplot
the two \nv profiles after appropriately scaling (see text) the \nvb 
assuming complete coverage. The good matching of profiles 
suggests that \nv absorption line covers the background source
completely.}
\label{fig3}
\end{figure}

\begin{figure}
\centerline{\vbox{
\psfig{figure=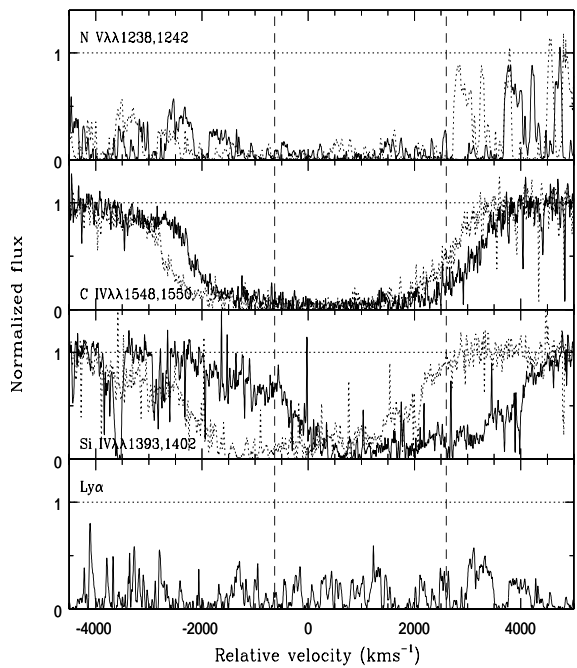,height=9.5cm,width=8.0cm,angle=0}
}}
\caption[]{Velocity plots for the blue component centered at $z_{abs}$~=~4.685.
Second member of the doublets is overlaid as a dotted line.
Vertical dashed lines mark the velocity range over which \siiv absorption line 
is seen. Note that \civ absorption is seen over a larger velocity range.
The non-zero residual flux at the expected position of \lya absorption
is used for obtaining the upper limit on the \hi column density.
}
\label{fig4}
\end{figure}
\subsection {The red component at $z_{\rm abs}$~=~4.855}
The absorption trough in this component is $\sim2000$ \kms~ wide. 
The absorption profiles of different transitions centered at 
$z_{abs}=4.855$ are shown in Fig.~\ref{fig2}. The absorption profile 
of the \siiv doublet that is comparatively less saturated than that of
the \civ doublet suggests the presence of 2 relatively narrow components 
denoted by l1 and l2 and 3 broad subcomponents denoted by a, b and c. 
While \lya and \civ are detected in  all these components, 
\nv absorption is {\bf detected} only in the broad subcomponents (see Fig.~\ref{fig2}).

\par\noindent
%
%
%
All the obviously saturated absorption lines have zero residual
(C~{\sc iv} and Lyman-$\alpha$, see Fig.~\ref{fig2}). 
The covering factor of the gas is therefore unity.
In case of complete coverage the observed residual intensities R$_1$
and R$_2$ in the first and second member of the doublet are related as
R$_1$=R$_2^2$ (see Srianand and Shankarnaraynan 1999).
%
%
%
%
Indeed, even though part of the \nv absorption line is contaminated by  
atmospheric absorption (see Fig.~\ref{fig3}), the velocity ranges in the 
profiles that are free of blending are consistent with complete coverage
(see top panel in Fig.~\ref{fig3}). As emission in this wavelength range
is contributed by both continuum and broad emission lines the corresponding 
absorbing gas has to be located outside the broad line region (BLR). As the 
\nv lines are weak we can derive the N~{\sc v} column densities. However, 
due to saturation effects, we can determine only lower limits for 
\civ and \hi column densities. 
It can be seen on Fig.~\ref{fig2} that Si~{\sc iv}$\lambda$1402 is only 
partly saturated which means that, although very uncertain,
we can estimate the corresponding column density.
We use also the Si~{\sc iv}$\lambda$1402 profile as
a template to determine upper limits on the column densities of 
non detected species (see below).
\par\noindent
The column density in each velocity pixel 
was obtained using the relation 
\begin{equation}
 N(v)=3.768\times 10^{+14}~\tau/f\lambda~{\rm cm^{-2}~km^{-1}s}
\end{equation}
where, $\tau$, $f$ and $\lambda$ are the optical depth, oscillator
strength and rest wavelength respectively. The total column
density is obtained by integrating $N(v)$ over the velocity range
$-$1120 to 1160 \kms. The results are presented in Table~\ref{tab1}.
Rest wavelengths and oscillator strengths used here are taken from  
Verner et al. (1994).
The two lines of the C~{\sc iv} doublet are blended and we use 
an effective oscillator strength, $f=f_1+f_2$, and mean
wavelength, ($\lambda_1+\lambda_2$)/2, to compute the lower
limit on the column density.
In the case of N~{\sc v}, we use the \nva line for component a and 
\nvb line for components b and c that are free of blending. 
Contamination by atmospheric features (see Fig.~\ref{fig3}) is corrected 
by carefully masking the atmospheric contamination using the normalised
spectrum of Q~1122$-$1628.
\par\noindent
Singly ionized species such as Si~{\sc ii}, C~{\sc ii} and Al~{\sc ii}
are absent and the spectrum at the expected position of Al~{\sc iii} lines is 
unfortunately very noisy. As said above, we evaluate upper limits for 
column densities of the latter species using the  \siivb profile (which is 
not completely saturated and least 
affected by atmospheric contamination) extending from $-$1120 to 1160 \kms~
as a template. The scaling factor
$k$~=~[$Nf\lambda$]$_{X^+}$/[$Nf\lambda$]$_{template}$ between the two optical depths 
$\tau$$_{template}$ and $\tau$$_{X^+}$ for species $X^+$ is then obtained 
by minimizing
\begin{equation}
\alpha = \Sigma(\tau_{\rm X^+} - k*\tau_{template})^{2}
\end{equation}
For Si~{\sc ii} and C~{\sc ii}, $k$ comes out to be 0.031 and 0.035 
respectively. The corresponding upper limits on the column densities are given 
in Table~\ref{tab1}. Note that the error in the column density of \nv is 
mainly due to continuum placement uncertainties. We note that most of
the atmospheric absorption seen in the expected wavelength range have
consistent equivalent width. This suggests that our continuum fitting 
does not underpredict the absorption in the \nv region.
No flux is detected at the
expected position of Si~{\sc iii}$\lambda$1206 (see Fig.~\ref{fig1}). 
However, we could not use this to derive a lower limit on Si~{\sc iii} as
the same region is contaminated by a probable \nv absorption line from the blue
component (see Fig.~\ref{fig4}).

\par
\begin{figure}
\centerline{\vbox{
\psfig{figure=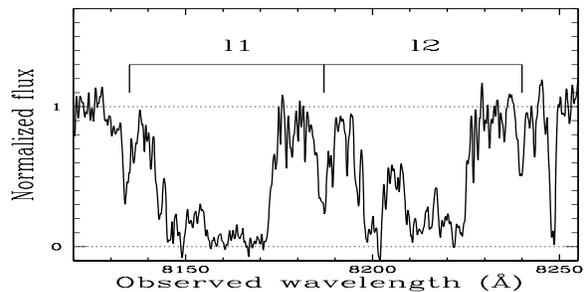,height=4.0cm,width=8.0cm,angle=0}
}}
\caption[]{Region of the spectrum containing the Si~{\sc iv} trough of the red 
BAL component at $z_{\rm abs}$~$\sim$~4.855. Note that systems l1 and l2 
have a velocity separation equal to that of the \siiv doublet splitting
suggesting the presence of line-locking in the system.}
\label{fig5}
\end{figure}

\noindent
Finally we notice that the velocity separation between the two narrow components 
l1 and l2 of the red component is very close to the Si~{\sc iv} doublet splitting.
This is shown in Fig.~\ref{fig5}. This is consistent with growing evidence for 
line-locked flows and radiative acceleration in BAL systems (see Srianand et al. 2002
and references therein).
\begin{table}
\caption{Column densities estimated using apparent optical
depth method with complete coverage. 
}
\begin{center}
\begin{tabular}{llc}
\hline
\hline
 \zabs    & species & \multicolumn{1} {c}{logN (cm$^{-2}$)}\\
\hline
4.855&\hi           & $\ge$15.67  \\
     &C~{\sc ii}    & $\le$14.21  \\
     &C~{\sc iv}    & $\ge$15.64  \\
     &Si~{\sc ii}   & $\le$14.13  \\
     &Si~{\sc iv}   & 15.33:       \\
     &N~{\sc v}     & 14.63$-$14.71\\
4.685& \hi          & $\le 16.00$ \\
     &\civ          & $\ge$15.94 \\ 
     &Si~{\sc ii}   & $\le$14.73 \\ 
     & Si~{\sc iv}  & $\ge$15.40 \\ 
\hline
\end{tabular}
\end{center}
\label{tab1}
\end{table}

\begin{table}
\caption{Chemical compositions considered in the models:}
\begin{center}
\begin{tabular}{ccccc}
\hline
\hline
      &Solar  &\multicolumn{3}{c} {Starburst*}\\
Metal &1.0 Z$_\odot$  & 1.0 Z$_{SB}$ & 0.5 Z$_{SB}$ & 0.20 Z$_{SB}$\\ 
\hline
C    &-3.44 &-3.90 &-4.22 & -4.63\\
N    &-4.03 &-4.60 &-5.34 & -6.51\\
O    &-3.13 &-3.00 &-3.30 & -3.70\\
Si   &-4.44 &-4.43 &-4.74 & -5.14\\
\hline
\end{tabular}
\label{abu}
\end{center}
* This form of Starburst abundances correspond to chemical evolution 
model M5a of Hamann \& Ferland {\bf (1993)}.
\end{table}

\subsection{The blue component at $z_{\rm abs}$~=~4.685}
The absorption profiles of different transitions centered at 
$z_{abs}=4.685$ are shown in Fig.~\ref{fig4}. Doublet partners for \civ
and \siiv are partially blended together. The non-zero residual flux for the 
blended \civab corresponds to a covering factor, $f_{\rm c}$ $\ge$0.90. 
It is interesting to note that the absorption is well detached from the
emission line profile. Thus the partial coverage reflects either
$\simeq 10\%$ contribution from the scattered light or partial
coverage of the continuum source.
Lower limits to column densities for the \siivab and \civab
were obtained assuming full coverage (see Table~\ref{tab1}).
Doing otherwise would only increase these limits.
Note that \civ absorption line extends over a larger velocity range as compared to 
\siiv absorption profile (see Fig.~\ref{fig4}). 
Like in the red component, singly ionized species \siii and \cii are
not detected which indicates that the neutral hydrogen column density
cannot be large (remember that metallicity in this kind of gas
is usually large). N~{\sc v} may be present but its redshifted position in 
the \lya forest overlaps with that of H~{\sc i} and Si~{\sc iii} from 
the red component (see Fig.~\ref{fig1}). The mean transmitted flux
at the \nv position is consistent with saturated absorption line.
\par\noindent
In the  Fig.~\ref{fig4} the intervening Lyman-$\alpha$ 
forest is clearly observed over the whole range corresponding to the BAL 
H~{\sc i} Lyman-$\alpha$ trough. This means that the H~{\sc i} optical depth 
cannot be large except if the covering factor of H~{\sc i} is much smaller than
that of C~{\sc iv}. This is not impossible (see e.g. Srianand et
al. 2002) but would be surprising given the non detection of
both C~{\sc ii} and Si~{\sc ii}.
By comparing the mean transmission in this region with that 
devoid of BAL absorption lines
we estimate that $\tau(v)$ for \hi  is less than 2. 
If we use the velocity range covered by the \civ profile 
($\simeq 4000$ \kms), we see that the \hi column density 
can not much larger than 10$^{16}$ cm$^{-2}$.
\begin{table*}
\caption{
Summary of the photoionization models considered for the red component.
}  
\begin{center}
\begin{tabular}{llclccc}
\hline
\hline
      \multicolumn{4}{c}{\bf Model parameters}     & \multicolumn {2}{c}{\bf allowed range in log U from}  & {\bf log(N(H(total)))*}\\ 
      Spectrum & enrichment& Metallicity(Z) & log N(H~{\sc i})*& ratios & observed N& \\
\hline
\\
      MF  & Solar &0.5Z$_{\odot}$,[Si/N]=3.0[Si/N]$_{\odot}$ & 17.00 & (-2.25,-2.15)  & $\approx$-2.2,-2.1& (19.99,20.11)\\
      
%


&&&&\\
  BAL &  "   & 1.0Z$_{\odot}$ & 16.00   & (-1.5,-1.1)          & $\approx$-1.1    &  20.00     \\
   "          &   "   & 1.5Z$_{\odot}$ &   "   & (-1.4,-0.95)         & $\approx$-1.0    &  19.90 \\
   
     &&&&&&\\ 
     
   "          &   "   & 0.05$_{\odot}$ & 17.00 & (-1.7,-0.75)         & $\approx$-0.7    & 21.42\\   
   "          &   "   & 0.1Z$_{\odot}$ &   "   & (-1.7,-0.7)          & (-0.9,-0.8)      & (21.53,21.62)\\
   "          &   "   & 0.2Z$_{\odot}$ &   "   & (-1.65,-0.65)        & $\approx$-1.0    & 21.36\\
   "          &   "   & 1.0Z$_{\odot}$ &   "   & (-1.4,-0.4)        & --                & --\\
   &&&&&&\\

   "    &  Starburst         & 1.0Z$_{SB}$    &   16.00  & (-1.4,-0.65)       & $\approx$-0.7    & 20.30\\
   "    &   "         & 5.0Z$_{SB}$    &   "   & (-1.4,-0.7)        & $\approx$(-1.0,-0.9)     & (19.32,19.41) \\
   "    &   "         & 9.0Z$_{SB}$   &   "   & (-1.15,-0.8)        & $\approx$-1.05   & 19.11 \\ 

   &&&&&&\\

   "    &   "         & 0.2Z$_{SB}$    & 17.00 & (-1.7,FP)          & $\approx$0.1     & 22.42 \\
   "    &   "         & 0.5Z$_{SB}$    &   "   & (-1.5,0.1)         & $\approx$(-0.6,-0.5)& 
   (21.60,21.70) \\
   "    &   "         & 1.0Z$_{SB}$    &   "   & (-1.3,0.1)         & $\approx$-0.8    & 21.12\\
   &&&&&& \\ \hline
\end{tabular} 
\end{center}
\label{tab3}
* in cm$^{-2}$
\end{table*}
\section{Discussion}
\begin{figure}
\centerline{\vbox{
\psfig{figure=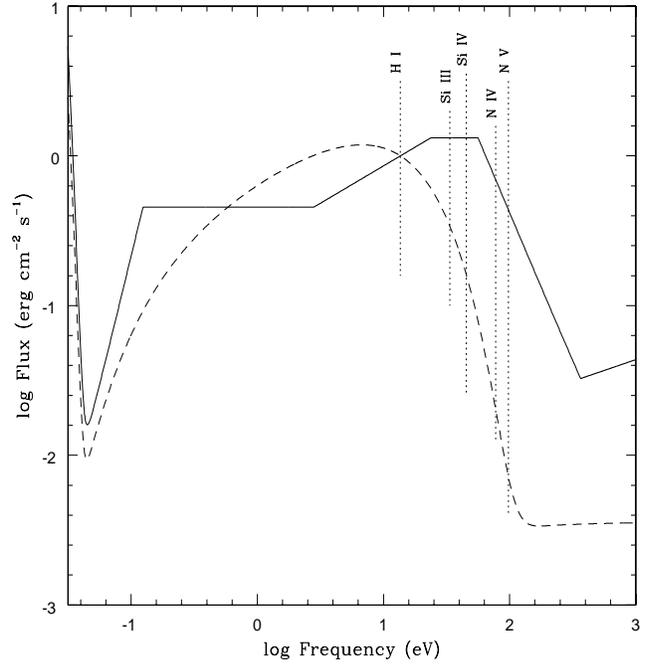,height=9.5cm,width=9.0cm,angle=0}
}}
\caption[]{ The spectral energy distribution (SED) of the ionizing 
radiation. The continuous and dashed curves correspond to the SED of
the MF and BAL spectrum respectively. The vertical dotted lines
give the ionization potential of the different species we are interested
in this study.
}
\label{fig6}
\end{figure}
\subsection {Physical conditions in the outflowing gas}

From the discussion in the previous Section, it is clear that
constraints are not strong enough to probe the physical conditions
in the blue component (\zabs~= 4.685). We therefore mainly concentrate on 
the red component (\zabs~= 4.855).
Recent Chandra observations (Vignali et al. 2001) fail to detect X-rays 
from the QSO SDSS J160501.21-011220.0. As discussed before,
the X-ray non-detection could be either due to large X-ray absorbing 
column density or a consequence of the QSO being intrinsically X-ray 
quiet. Indeed, there are already some reports that the X-ray continuum 
shapes of QSOs may evolve at $z\ge$2.5 (e.g. Vignali et al. 1999; Blair et al. 2000; 
Vignali et al. 2003). Assuming that the lack of X-rays in this source is 
due to absorption, Vignali et al. (2001) infer that the X-ray
non-detection for this object implies a total hydrogen column density 
of N(H(total))~$\ge 5.0 \times 10^{23}$~cm$^{-2}$, 
characteristic of low-ionization 
BAL~QSOs at low redshift. Note that the limit on the X-ray flux further 
implies a small optical to X-ray spectral index, $\alpha$$_{\rm ox}$~
$<$~$-$1.82, for this object. 
\par\noindent
The absence of singly ionized species \cii and \siii suggests that H~{\sc i} is 
optically thin at the Lyman limit. Thus, to probe the physical conditions in 
the red component, we run grids of photoionization models using Cloudy 
(Ferland 1996) in the range  log~$N$~(H~{\sc i})~(cm$^{-2}$)$\sim$~10$^{16}$ 
to 10$^{17}$ cm$^{-2}$, considering the gas being ionized by 
either an unattenuated QSO spectrum given by Mathews \& Ferland 
(1987) (hereafter MF spectrum; see Fig.~\ref{fig6}) or a modified
spectrum with little X-rays (hereafter BAL spectrum; 
see Fig.~\ref{fig6}) mimicking attenuation 
of a typical QSO spectrum by a large column density of ionized gas
(see below).
The models are run for different chemical composition,
either the Solar one or the so-called Starburst one (see Table~\ref{abu}).
In all the models, the calculations are stopped when the {\bf total}
neutral hydrogen column density reaches the limit we set (log $N$(H~{\sc i})~(cm$^{-2}$) 
= 16 or 17).
\begin{figure*}
\centerline{\vbox{
\psfig{figure=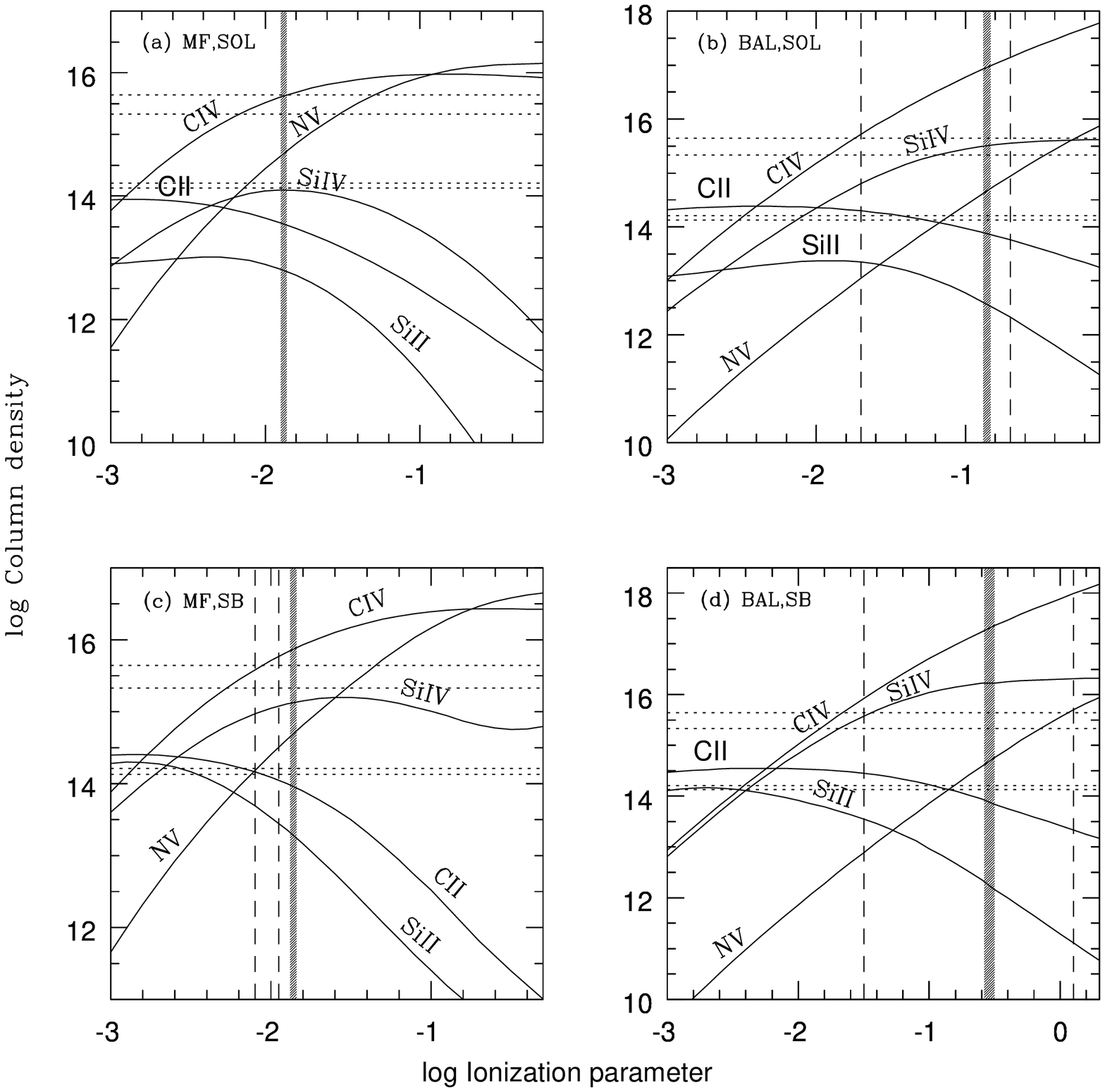,height=16cm,width=18.0cm,angle=0}
}}
\caption[]{Predicted column densities of different species as a function of
ionization parameter for log~$N$(H~{\sc i})~(cm$^{-2}$)~=~17.
Results given in panels (a) and (b) are obtained for the MF and BAL  spectrum with
metallicity 0.1 Z$_\odot$ and 0.5 Z$_\odot$ respectively. 
The results for MF and BAL ionizing spectrum with  metallicity 
0.1 Z$_{SB}$ of the Starburst enrichment (see Table~\ref{abu}) are given in panels (c) 
and (d) respectively. In each panel the horizontal dotted lines
(from bottom to top) mark
the limits on the column density of \siii, \cii, \siiv and \civ obtained
from our observations. While the former two are lower limits the latter
two are upper limits. The verticle dashed lines give the limits on
the ionization parameters based on the X~{\sc ii}/X~{\sc iv} ratios and
the observed ratios of \nv and \siiv. The verticle shaded region gives
the ionization parameter range over which the observed column density of
\nv is reproduced for the chosen nitrogen metallicity.}
\label{fig7}
\end{figure*}
\par\noindent
The input parameters of the models, that reproduce
the observed ratios as well as the column densities of individual species, 
are given in Table~\ref{tab3}. 
The corresponding total hydrogen column density 
in the allowed range of ionization parameters is given in the last
Column of the Table~\ref{tab3}. 
Results for some selected models with different ionizing spectra
and chemical composition for log~$N$(H~{\sc i})~(cm$^{-2})$)~=~17 
are given in Fig.~\ref{fig7}. In this figure we have plotted the column density
of different species as a function of ionization parameter. 
The horizontal dotted lines
mark limits on the observed  column densities of \siii, \siiv, \cii and
\civ. The minimum and maximum of the
allowed range for log $U$ are marked with verticle dashed lines. 
The former is obtained using either of the ratios
$N$(Si~{\sc ii})/$N$(Si~{\sc iv}) or $N$(C~{\sc ii})/N(C~{\sc iv})
and the latter is obtained from the observed $N$(N~{\sc v})/N(Si~{\sc iv})
ratio for metallicities assumed in the model. 
The range of ionization parameters for which the observed N~{\sc v} column 
density is reproduced is indicated as a verticle shaded region.
\par\noindent
We note that, if the \hi column 
density is close to the observed value (i.e. few 10$^{16}$ cm$^{-2}$) 
one needs overall metallicities larger than Solar to explain the 
observed column densities. 
We therefore mainly concentrate on models with $N$(H~{\sc i})$ = 10^{17}~{\rm cm}^{-2}$
that will give a conservative lower limit on the metal enrichment in the
system. 
\par\noindent
If the gas is ionized by a MF type spectrum and has a Solar composition, 
it can be seen from panel (a) in Fig.~\ref{fig7} that the observed constraints on the
ion ratios and the \nv and \civ column densities can be consistently
reproduced for log U $\simeq-1.9$ (see the shaded region). However the 
predicted column density of \siiv is lower than the observed value by more 
than an order of magnitude. It is clear that the observed value
can be reproduced only if [Si/H]$\ge$[Si/H]$_\odot$.
Increasing the overall abundance actually reduces the required overabundance of
Si with respect to N. For example if one assumes  Z~=~0.5~Z$_\odot$ 
the observed column densities can be reproduced consistently with lower
ionization parameter and a [Si/N] ratio higher than Solar by a factor of 3 
only (see Table.~\ref{tab3}). Larger metallicities are not possible however 
as C~{\sc ii} would then be detectable.
\par\noindent
The enhancement of $N$(Si~{\sc iv}) with respect to $N$(N~{\sc v}) can  
be achieved by either enhancing the silicon metallicity with respect to 
the nitrogen one as discussed above or by suppressing the N~{\sc iv} ionizing photons. 
The first possibility can be achieved in a chemical enrichment model where 
the secondary production of N has not yet begun to dominate. 
The latter possibility naturally arise if 
the cloud sees an attenuated X-ray spectrum, similar to what we observe 
directly for SDSS~J160501.21$-$011220.0. We explore this possibility in the following.
\par\noindent
We model the absorbed spectrum (called BAL spectrum in the following) as the composite of a 
black body spectrum with temperature $T$ =~150,000 K, plus a power-law 
spectrum with $\alpha_{uv}$ =~$-$0.5 and $\alpha_{x}$ =~$-$1.0 and a relative scaling 
$\alpha$$_{ox}$~=~$-$2.0.
From Fig.~\ref{fig6}, it is clear that for a given number of hydrogen ionizing 
photons this spectrum has less photons to ionize Si~{\sc iii} and N~{\sc iv}
compared to the MF spectrum. 
Thus a given ionization state will be produced by higher values of log U
and hence higher total hydrogen column densities. However the major difference
is the presence of a soft X-ray excess in the MF spectrum that implies
more N~{\sc iv} ionizing photons for a given number of Si~{\sc iii} ionizing
photons compared to the BAL spectrum. 
For a column density of log~$N$~(H~{\sc i})~(cm$^{-2}$)$\approx$17.00, observations can be
reproduced by models with metallicities 0.05Z$_{\odot}$$\le$Z$\le$0.2Z$_{\odot}$ 
(Table~\ref{tab3}). This is also demonstrated in panel (b) of Fig.~\ref{fig7}.
As expected, in these models the results are consistent with observations 
at an ionization parameter an order of magnitude higher than that required 
for the MF ionizing spectrum. Thus, 
we can reproduce the observed column densities with Sub-Solar metallicity 
and Solar abundance ratios. 
\par\noindent
It is usual 
in the case of the BLR clouds to explain an enhanced nitrogen abundance
by rapid star formation nucleosynthesis.
We consider such a chemical enrichment in our models
using option "starburst abundances" in Cloudy (see Table~\ref{abu}). 
The details of the models are given in Hamann \& Ferland (1993). These models
have problems in reproducing the observations 
when a MF ionizing spectrum is considered (see panel
(c) in Fig.~\ref{fig7}). 
However, the observed ratios are reproduced if the BAL spectrum is
used instead
(see Table~\ref{tab3} and panel (d) in Fig.~\ref{fig7}) for overall
metallicity  $\ge 0.2$ and log~{N(H~{\sc i})}~(cm$^{-2}$) in the range 16 to 17.
\par\noindent
It is interesting to note that for all the models considered here the 
total hydrogen column density is at least an order of
magnitude less than the amount required (i.e. $N$(H(total)) $\ge5\times 10^{23}$ 
cm$^{-2}$) to explain the lack of X-rays in this QSO by absorption
of a standard QSO spectrum (Table~\ref{tab3}). We also run models
by fixing the total hydrogen column 
density to be $N$(H(total)) $\ge5\times 10^{23}$ cm$^{-2}$.
For MF ionizing spectrum we do not find any model that can
reproduce the constraints noted in Table~{\ref{tab1}}.
%
%
Thus, we can conclude that either SDSS J160501.21$-$011220.0 is intrinsically X-ray weak or,
if there is any Compton thick screen between the quasar and us, it has nothing 
to do with the gas that is associated with the red component. 
\par\noindent
The similar characteristics of the blue component together with
the low $N$(H~{\sc i}) column density of the system strongly suggests
that the blue component as well is not Compton thick.
\subsection{Formation redshift of the blackhole} 
We estimate the mass of the black hole from the Eddington accretion. 
For this we estimate the rest frame luminosity at 2500~\AA~ using the 
low dispersion spectra obtained by the SDSS group (Fan et al. 2000)
to be L$_\nu$$=$1.37$\times$10$^{31}$ erg~s$^{-1}$~Hz$^{-1}$.
Applying bolometric correction 
as suggested by Elvis et al. (1994), we obtain  a bolometric luminosity of
L$_{bol}$=9.21$\times$10$^{46}$ erg~s$^{-1}$.
Assuming Eddington accretion  
we derive the black hole mass, 
$M_{\rm BH}$~=~7.1$\times$10$^{8}$$\eta^{-1}$~M$_{\odot}$
(see  Haiman \& Loeb 2001).
Here, $\eta$ is the efficiency parameter (i.e. accretion
luminosity given in the units of Eddington luminosity).
If $\epsilon$=L$_{bol}$/\.M$_{BH}$c$^2$ is the radiative 
efficiency for a mass accretion rate \.M$_{BH}$ then
the natural e-folding time scale for the growth of a single seed
can be written as 
\begin{equation}
t=M_{BH}/$\.M$_{BH}=4.0\times10^7\epsilon_{0.1}\eta^{-1} yr
\end{equation}
It will therefore take about ln(7.1$\times$10$^{8}$$\eta^{-1}$M$_{\odot}$/10~M$_{\odot}$)
= 18.1 e-folding times  (or t $\ge 0.7\epsilon_{0.1}\eta^{-1}$ Gyr) 
for the black hole to grow to the above estimated mass from a
stellar-mass seed of 10~M$_\odot$. For the assumed value of $\epsilon$~
$\approx$~0.1 and $\eta$~$\approx$~1.0  
this corresponds to a formation epoch of the 
black hole close to $z \approx11$ for the cosmological model considered
here.

\section {Summary}
We have presented the analysis of broad absorption lines (BALs) 
seen in the spectrum of the \zem $\simeq$4.92 QSO SDSS J160501.21-011220.0. 
Our high spectral resolution UVES spectrum shows two well detached 
absorption line systems at \zabs= 4.685 and 4.855. The system at \zabs= 4.855 
covers the background source completely suggesting that the gas is 
located outside the broad emission line region. On the contrary the 
system at \zabs= 4.685, which is redshifted on top of the quasar 
continuum, has a covering factor of the order of 0.9. Physical conditions 
are investigated in the BAL system at \zabs= 4.855 using detailed 
photoionization models.
The observed \hi column density and the limits on \cii and \siii
absorptions suggest that log $N$(H~{\sc i})~(cm$^{-2}$) is in the range 16--17.
The observed column densities of \nv, \siiv and \civ
in the \zabs = 4.855 component require, 
unlike what is derived when analysing broad emission lines, that 
nitrogen is underabundant by more than a factor of 3 compared to 
silicon if the gas is photoionized by a typical QSO spectrum. 
Thus, if the gas is ionized by a standard 
MF spectrum, the chemical enrichment of the cloud is
different from that required by emission line clouds. 
We show however that the relative supression in the \nv column densities 
can be reproduced for Solar abundance ratios or abundance ratios typical of
rapid Starburst nucleosynthesis if we use an ionizing spectrum 
that is devoid of X-rays. 
Thus if the composition of BAL is like that of
the emission line regions it is most likely that the cloud 
sees an ionizing spectrum similar to what we observe from this QSO
that is strongly attenuated in the X-rays. This is 
consistent with the fact that none of our models have high 
enough Compton optical depth to be able to remove X-rays from the QSO.
Similar arguments lead to the conclusion that the system at \zabs = 4.685
as well is not Compton thick. 
\par\noindent
Using simple Eddington arguments we show that the mass of the central
black hole is $\sim 8\times 10^8$ M$_\odot$. This suggests that the 
accretion onto a seed black hole must have started as early as
$z$ $\sim$ 11. This gives a typical formation epoch for the host galaxy
of the QSO.
%
\section*{acknowledgments}
This work was supported in part by the European Communities RTN network
"The Physics of the Intergalactic Medium".  We wish to thank
Dr. Fan for making the low dispersion data on SDSS J160501.21$-$011220.0
available to us.
%
%
%

\end{document}